\documentclass[prb,twocolumn,showpacs,floatfix,preprintnumbers,amsmath,amssymb]{revtex4}
\usepackage{epsfig}
\usepackage[latin9]{inputenc}
\usepackage{color}
\usepackage{hyperref}
\usepackage{graphicx}
\usepackage{sidecap}
\usepackage{hypernat}
\begin{document}

\title{Physical Properties of Ba$_\mathrm{2}$Mn$_\mathrm{2}$Sb$_\mathrm{2}$O Single Crystals}

\author{J. Li,$^1$ C. E. Ekuma,$^{1,2}$ I. Vekhter,$^1$ M. Jarrell,$^{1,2}$ J. Moreno,$^{1,2}$
S. Stadler,$^1$ A. B. Karki,$^1$ and R. Jin$^1$}
\affiliation{$^1$Department of Physics and Astronomy,
Louisiana State University,
Baton Rouge, LA 70803, USA\\
$^2$Center for Computation and Technology,
Louisiana State University,
Baton Rouge, LA 70803, USA}

\date{\today}

\begin{abstract}
\noindent We report both experimental and theoretical investigations of the physical properties of
Ba$_\mathrm{2}$Mn$_\mathrm{2}$Sb$_\mathrm{2}$O single crystals.
This material exhibits a hexagonal structure
with lattice constants: a = 4.7029(15) \AA{} and c = 19.9401(27) \AA{}, as obtained from powder
X-ray diffraction measurements, and in agreement with structural optimization through density functional theory 
(DFT) calculations.
The magnetic susceptibility and specific heat show anomalies at
T$_\mathrm{N}$ = 60 K, consistent with antiferromagnetic ordering. However,
the magnitude of T$_\mathrm{N}$ is significantly smaller than the Curie-Weiss
temperature ($\mid$$\mathrm{\Theta_{CW}}$$\mid$ $\approx$ 560 K), suggesting a magnetic
system of reduced dimensionality.  The temperature dependence of both the in-plane and out-of-plane resistivity
changes from an activated at $T$ $>$ T$_\mathrm{x}$ $\sim$ 200 K
to a logarithmic at $T$ $<$ T$_\mathrm{x}$. Correspondingly, the magnetic
susceptibility displays a bump at T$_\mathrm{x}$. DFT calculations at the DFT + U level support 
the experimental observation of an
antiferromagnetic ground state.
\end{abstract}


\maketitle

\section{Introduction}
The discovery of superconductivity in Fe-based layered compounds has sparked immense interest
in the physics and chemistry communities, reminiscent of the excitement that accompanied
the discovery of the high-T$_\mathrm{c}$ cuprate superconductors more than two decades
ago.\cite{Kamihara2006,Bednorz1986} Although both families form similar layered structures,
the building blocks are different, with tetrahedral FeX$_{4}$ (X = As, Se) for
Fe-based superconductors but octahedral CuO$_{6}$ in Cu-based superconductors. In addition,
Fe can be partially or even completely replaced by other transition
metals,\cite{Sefat2008,Sefat2009,PhysRevB.85.144513} while doping on the Cu site of cuprates
immediately kills superconductivity. \cite{PhysRevB.65.172501}
This suggests that superconductivity in materials with the tetrahedral building block is more
tolerant to various dopings. \cite{Sefat2008,Sefat2009,PhysRevB.85.144513,PhysRevLett.104.057006,
PhysRevB.82.024519,PhysRevB.83.144516,PhysRevB.83.224505}
In particular, magnetic elements such as Co and
Ni induce extremely rich phase diagrams including the coexistence of superconductivity and
magnetism. \cite{PhysRevLett.104.057006,PhysRevB.82.024519,PhysRevB.83.144516,PhysRevB.83.224505}
In the search for new superconductors
with structures similar to the Fe-based materials, but without Fe, BaNi$_{2}$As$_{2}$
was found to superconduct at 0.7 K. \cite{Sefat2009} So far, there is no evidence of superconductivity in Mn-based compounds.

Many Mn-based compounds form layered structures and order magnetically. For example, BaMn$_{2}$As$_{2}$
shares the same structure with BaFe$_{2}$As$_{2}$, one of the parent compounds of Fe-based superconductors.
Although it orders antiferromagnetically below 625 K,\cite{PhysRevB.80.100403} its spin structure is
different from that of BaFe$_{2}$As$_{2}$. It is also electrically insulating due to strong spin dependent
Mn-As hybridization.\cite{PhysRevB.79.075120} Recently, another layered compound, Sr$_{2}$Mn$_{3}$As$_{2}$O$_{2}$,
has been found to exhibit a spin-glass transition due to competing ferromagnetic and antiferromagnetic
interactions.\cite{Nath2010} While it crystallizes in a tetragonal structure, Mn occupies two different
environments, one in the MnAs$_{4}$ tetrahedra and the other in the MnO$_{2}$ sheets.\cite{Brechtel1981,Ozawa2008}
Other compounds with the same structure such as A$_{2}$Mn$_{3}$Pn$_{2}$O$_{2}$ (A = Sr, Ba; Pn = P, As, Sb, Bi) have
been synthesized.\cite{Stetson1991,Brechtel1979,Nath2010} The magnetic moment due to the Mn in edge-shared
MnPn$_{4}$ orders antiferromagnetically at low temperatures. This motivates us to study
Ba$_\mathrm{2}$Mn$_\mathrm{2}$Sb$_\mathrm{2}$O, which has a double layered MnSb$_{3}$O tetrahedra
edge shared in the $\textit{ab}$-plane, but corner shared along the $\textit{c}$-direction (see
Fig.~\ref{fig:BaMnSbO_crystal}(a)). All Mn sites are equivalent.

While it has existed since 1981,\cite{Brechtel1981} Ba$_\mathrm{2}$Mn$_\mathrm{2}$Sb$_\mathrm{2}$O is 
only known to exhibit a hexagonal structure with space group P6$_3$/mmc.  Its physical properties
have not yet been investigated. Here, we report the crystal growth,
structural, electronic, magnetic, and
thermodynamic properties of Ba$_\mathrm{2}$Mn$_\mathrm{2}$Sb$_\mathrm{2}$O. Computational
studies have also been carried out to help understand its physical properties.

\section{Experimental and Computational Details}
\label{sec:experiment}
\begin{figure}
\centering
\includegraphics*[trim = 9mm 30mm 10mm 10mm, clip,totalheight=0.18\textheight, width=3.2in]{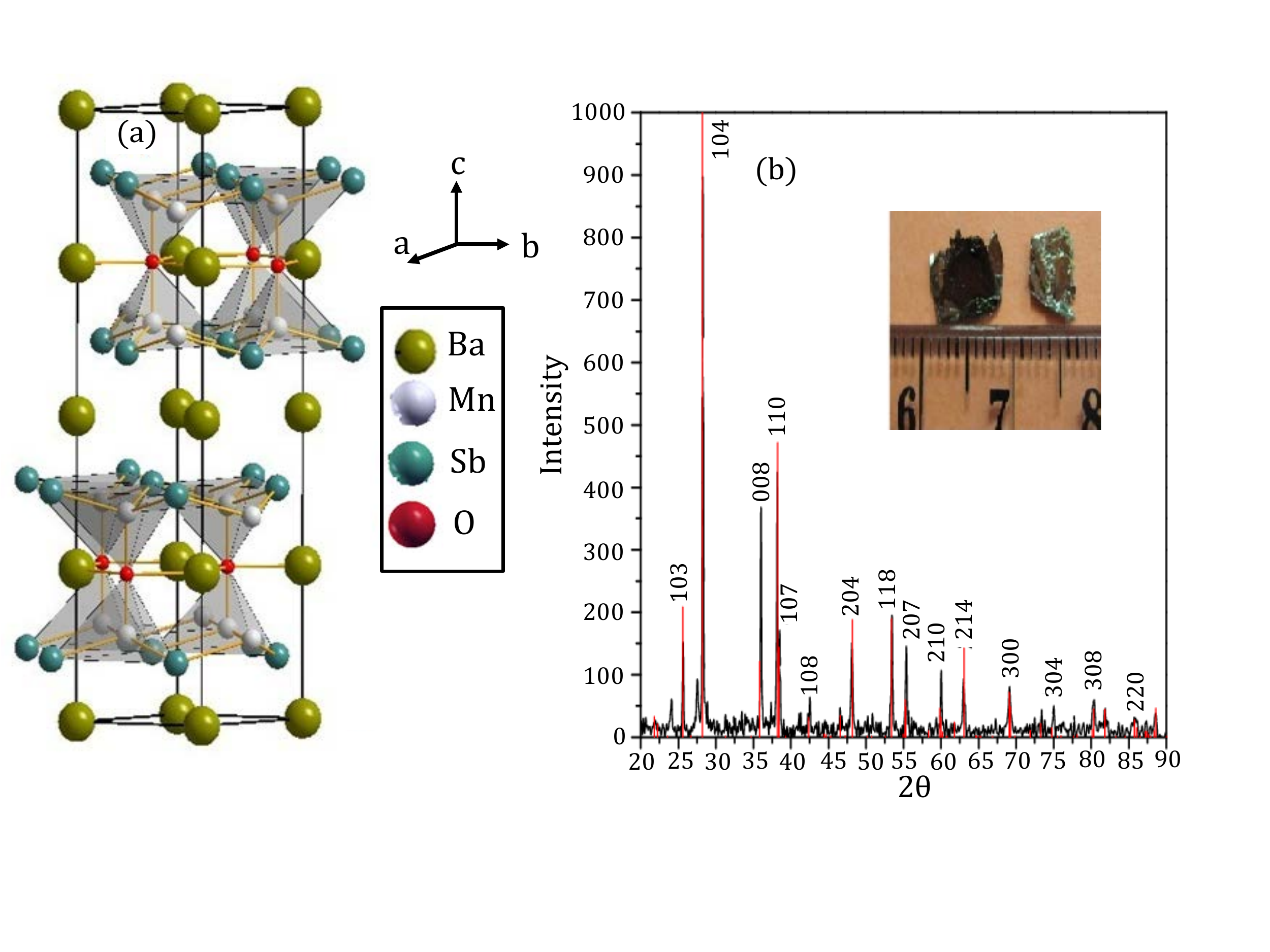}
\caption{(Color online) (a) Crystal structure of the layered Ba$_\mathrm{2}$Mn$_\mathrm{2}$Sb$_\mathrm{2}$O;
(b) X-ray powder diffraction pattern at room temperature for single crystal
Ba$_\mathrm{2}$Mn$_\mathrm{2}$Sb$_\mathrm{2}$O (black line), which matches well with the standard X-ray powder
diffraction pattern (red line).  The inset shows
single crystal Ba$_\mathrm{2}$Mn$_\mathrm{2}$Sb$_\mathrm{2}$O with typical size of 5 mm $\times$ 5 mm $\times$ 0.2 mm.}
\label{fig:BaMnSbO_crystal}
\end{figure}

To grow Ba$_\mathrm{2}$Mn$_\mathrm{2}$Sb$_\mathrm{2}$O single crystals, stoichiometric amounts of high-purity
Ba pieces (99 \% Alfa Aesar), Mn powder (99.95 \% Alfa Aesar), Sb powder (99.5 \% Alfa Aesar) and MnO$_{2}$ powder
(99.997 \% Alfa Aesar) are mixed in the ratio 4 : 3 : 4 : 1. The mixture is placed in an alumina crucible 
sealed in an evacuated quartz tube. The samples are first heated to 1150$^{o}$C at the rate of 150$^{o}$C/h. This
temperature is held for 15 h, after which the samples are cooled to 700$^{o}$C at the rate of 4$^{o}$C/h, and finally cooled down to
room temperature by turning off the power. Shiny black plate-like crystals are obtained 
without requiring additional process. These crystals have a typical size of 5
$\times$ 5 $\times$ 0.2 mm$^{3}$, as shown in the inset of Fig.~\ref{fig:BaMnSbO_crystal}(b).

The phase of the as-grown crystals was characterized using a Scintag XDS-2000 powder X-ray diffractometer using
Cu K$\alpha$ radiation ($\lambda$ = 1.54056 \AA{}). Electrical transport and heat capacity measurements
were carried out with a Quantum Design physical property measurement system (PPMS). 
Both in-plane and c-axis resistivities were measured using the standard four-probe technique. 
Thin Pt wires were used
as electrical leads and attached to the sample using silver paste. The magnetic properties were measured using
a Quantum Design magnetic property measurement system (MPMS) between 2 and 400 K, and a vibrating sample
magnetometer (VSM) between 300 K and 700 K in a PPMS.

We compare the experimental results with the electronic structure calculations. These calculations were performed using the general potential linearized augmented plane wave (LAPW)
method \cite{Singh2006} as implemented in the WIEN2K electronic structure code.\cite{Blaha2001} LAPW
radii of 2.5 Bohr were used for Ba and Sb, while LAPW radii of 2.01 Bohr and 1.78 Bohr were used for Mn
and O, respectively. The calculations were done for the hexagonal crystal structure (space group
P6$_{3}$/mmc) with optimized lattice parameters of a = 4.7096 \AA{} and c = 20.0298 \AA{} obtained
from our computations.  The unit cell of Ba$_\mathrm{2}$Mn$_\mathrm{2}$Sb$_\mathrm{2}$O contains
two inequivalent Ba sites. Table~\ref{table:BaMnSbO_Wyck} shows the locations of these inequivalent Ba sites as well as those of other atoms in the crystal with their Wyckoff positions. The computations
were based on RK$_{max}$ = 7.0, where $R$ is the smallest LAPW sphere radius and K$_{max}$ is the
interstitial plane wave cutoff. The Brillouin zone integration was done using a 9 $\times$ 9 $\times$ 2
reciprocal space mesh. The core states were treated relativistically, and the spin-orbit interaction was
included self-consistently through the second variational step. In all our computations, no shape
approximation was made on either the potential or the charge density.

\begin{table*}[ht]
\caption{The atomic positions and the Wyckoff numbers for the atoms in Ba$_\mathrm{2}$Mn$_\mathrm{2}$Sb$_\mathrm{2}$O.
Experimental data is adapted from Ref. 14.}   
\centering                          
\begin{tabular}{ c | c | c | c | c | c | c | c}            
\hline\hline                        
 Site & Wyckoff & Symmetry & Charge & x & y & z & Atomic environment \\   
\hline     

\multicolumn{8}{c}{Experimental atomic positions}  \\
\hline
Mn1 & 4f & 3m. & Mn2+ & $\frac{1}{3}$ & $\frac{2}{3}$ & 0.6495 & single atom O\\
Sb2& 4f & 3m. & Sb3- & $\frac{1}{3}$ & $\frac{2}{3}$ & 0.3873 & non-coplanar triangle Mn$_{3}$\\
O1& 2d & -6m2 & O2- & $\frac{1}{3}$ & $\frac{2}{3}$ & $\frac{3}{4}$ & trigonal bipyramid Mn$_{2}$Ba$_{3}$\\
Ba4& 2b & -6m2 & Ba2+ & 0 & 0 & $\frac{1}{4}$ & coplanar triangle O$_{3}$\\
Ba5& 2a & -3m & Ba2+ & 0 & 0 & 0 & octahedron Sb$_{6}$\\
\hline    
\multicolumn{8}{c}{Atomic positions as obtained from geometrical optimization}  \\
\hline
Mn1 & 4f & 3m. & Mn2+ & $\frac{1}{3}$ & $\frac{2}{3}$ & 0.6506 & single atom O\\               
Sb2& 4f & 3m. & Sb3- & $\frac{1}{3}$ & $\frac{2}{3}$ & 0.39467 & non-coplanar triangle Mn$_{3}$\\
O1& 2d & -6m2 & O2- & $\frac{1}{3}$ & $\frac{2}{3}$ & $\frac{3}{4}$ & trigonal bipyramid Mn$_{2}$Ba$_{3}$\\
Ba4& 2b & -6m2 & Ba2+ & 0 & 0 & $\frac{1}{4}$ & coplanar triangle O$_{3}$\\
Ba5& 2a & -3m & Ba2+ & 0 & 0 & 0 & octahedron Sb$_{6}$\\
\hline\hline
\end{tabular}
\label{table:BaMnSbO_Wyck}          
\end{table*}

One difficulty in the computation of material properties is that the band gaps and related
properties of most materials are generally underestimated by the standard density functional
theory (DFT) approximations.  To avoid this, we utilized the constrained DFT + U scheme
of \citet{Anisimov1991} as implemented by \citet{Madsen2005} in WIEN2K. With this approach,
the effective Coulomb interaction (U$_{eff}$) on the Mn $\textit{d}$ state is calculated self-consistently.
The computed value of U$_{eff}$ is 7.07 eV. The DFT part of the computation
utilized the Perdew-Burke-Ernzerhof generalized gradient approximation (PBE-GGA).\cite{Perdew1996}

\section{Results and Discussion}
\label{sec:results}
The X-ray powder diffraction measurements were performed at room temperature by crushing as-grown single
crystals. Figure~\ref{fig:BaMnSbO_crystal}(b) shows the X-ray diffraction pattern of
Ba$_\mathrm{2}$Mn$_\mathrm{2}$Sb$_\mathrm{2}$O. As indicated in the figure, all peaks can be
indexed to the hexagonal structure with space group P6$_{3}$/mmc as previously
reported.\cite{Brechtel1981} From our X-ray diffraction data, we obtain the lattice
parameters: a = 4.7029 \AA{} and c = 19.9401 \AA{} at room temperature, in agreement with
that reported by \citet{Brechtel1981}

\begin{SCfigure*}
\centering
\raisebox{0cm}{\includegraphics*[trim = 18mm 10mm 8mm 0mm, totalheight=0.39\textheight, width=1.4\columnwidth,clip=true]{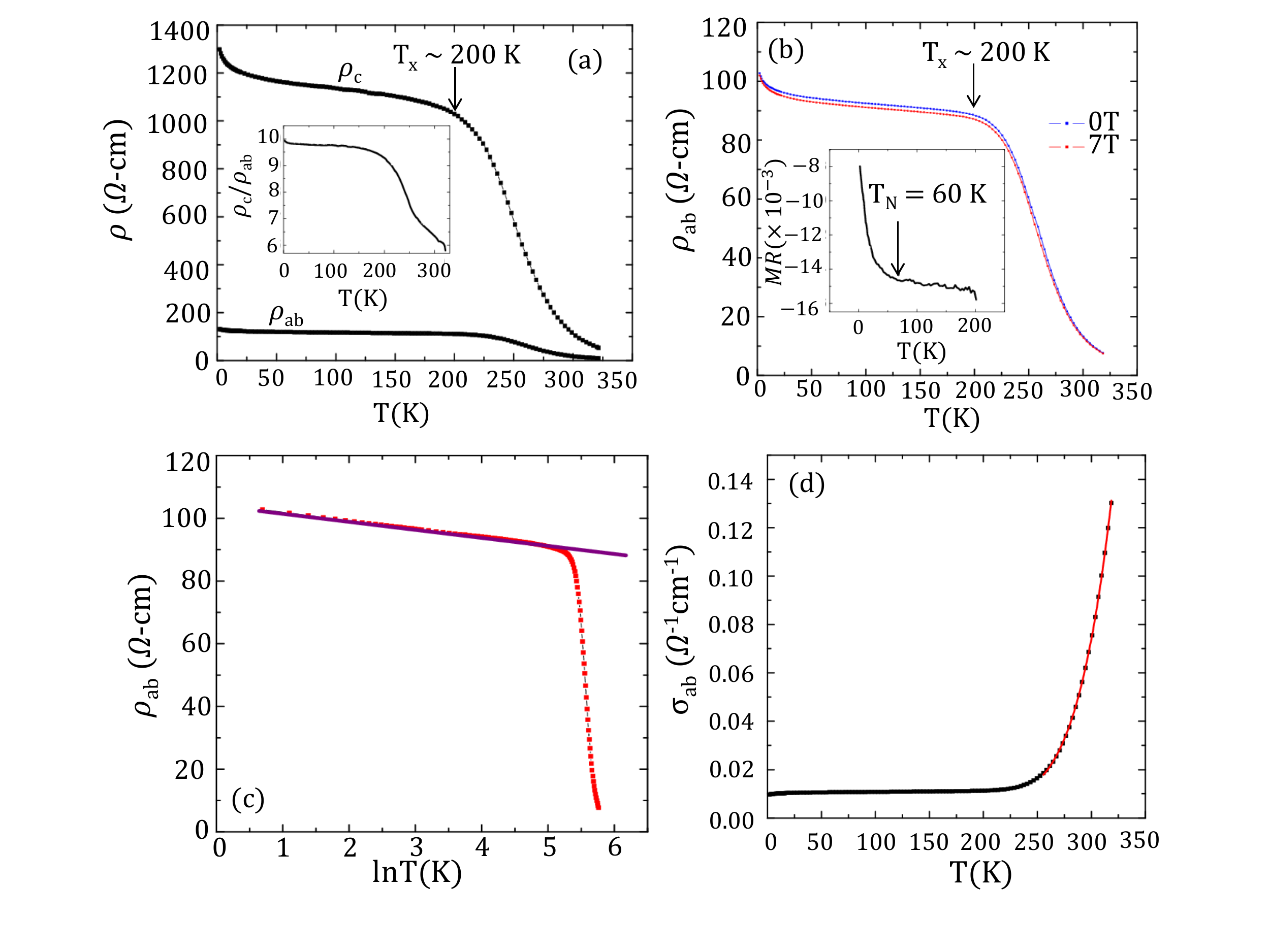}}
\caption{(Color online) (a) In-plane ($\rho_{ab}$) and out-of-plane ($\rho_{c}$)  resistivities of
Ba$_\mathrm{2}$Mn$_\mathrm{2}$Sb$_\mathrm{2}$O single crystal. The inset shows the
temperature dependence of resistivity anisotropy ($\rho_{c}$/$\rho_{ab}$); (b) Temperature dependence of
$\rho_{ab}$ under zero and 7 $T$ applied perpendicular to the $ab$-plane. The inset shows the temperature dependence
of magnetoresistivity MR (see the text); (c) In-plane resistivity plotted as $\rho_{ab}$ versus ln $T$.
The solid line is the fit of $\rho_{ab}$ below 200 K (see the text); (d) Temperature dependence of the in-plane conductivity
($\sigma_{ab}$). The red line is the fit of $\sigma_{ab}$ above 250 K with $\sigma_{ab}$ $=$ $\sigma_{0}$ $+$ $B$
$exp(- \frac{\Delta}{2 k_{B} T})$.}
\label{fig:BaMnSbO_resistivity}
\end{SCfigure*}

Figure~\ref{fig:BaMnSbO_resistivity}(a) displays the temperature dependence of in-plane
($\rho_{ab}$) and out-of-plane ($\rho_{c}$) resistivities of Ba$_\mathrm{2}$Mn$_\mathrm{2}$Sb$_\mathrm{2}$O
between 2 and 320 K. Note both $\rho_{ab}$ and $\rho_{c}$ increase with decreasing temperature.
However, the change is much slower below T$_\mathrm{x}$ $\sim$ 200 K than it is at high temperatures.
Correspondingly, the anisotropy, $\rho_{c}$/$\rho_{ab}$, is almost constant below T$_\mathrm{x}$ as shown in
the inset of Fig.~\ref{fig:BaMnSbO_resistivity}(a), and is comparable to that seen in the 
AFe$_\mathrm{2}$As$_\mathrm{2}$ (A = Ba, Sr, Ca)
system.\cite{Tanatar2009} As shown in Fig.~\ref{fig:BaMnSbO_resistivity}(b), the application of a magnetic
field, $H$ $=$ 7 T, normal to the $\textit{ab}$ plane (H $\perp$ $\textit{ab}$) results in a small reduction
of $\rho_{ab}$. This means that the magnetoresistance,
$MR$ $=$ $\displaystyle \frac{\rho_{ab}(H)- \rho_{ab}(0)}{\rho_{ab}(0)}$,
is negative, as shown in the inset of Fig.~\ref{fig:BaMnSbO_resistivity}(b). Note the magnitude of $MR$ decreases
more rapidly below $\sim$ 60 K.

In analyzing the temperature dependence of the electrical resistivity, we find that both $\rho_{ab}$(T) and
$\rho_{c}$(T) exhibit quantitatively similar behaviors.
Figure~\ref{fig:BaMnSbO_resistivity}(c) shows $\rho_{ab}$ versus $\ln$$T$, which reveals a logarithmic dependence
below
T$_\mathrm{x}$. From 2 to 200 K we obtain $\rho_{ab}$ = $-2.64 ln T + 104.6$ $\Omega$-cm, plotted as a solid line
in Fig.~\ref{fig:BaMnSbO_resistivity}(c). Above T$_\mathrm{x}$ the resistivity decreases exponentially, so that
we fit
the in-plane electrical conductivity data as shown in Fig.~\ref{fig:BaMnSbO_resistivity}(d)
using the formula for the conductivity, $\sigma_{ab}$ $=$ $\sigma_{0}$+$B$ $exp(- \frac{\Delta}{2 k_{B} T})$
($\sigma_{0}$ and B are constants, $k_B$ is the Boltzmann constant,
and $\Delta$ is the activation energy).  The fit (see the solid line in
Fig.~\ref{fig:BaMnSbO_resistivity}(d))
yields $\sigma_{0}$ $=$ 0.086 $\Omega^{-1}$$cm^{-1}$, B $=$
2095 $\Omega^{-1}$$cm^{-1}$, and $\Delta$ $\sim$ 0.59 eV.
A similar gap value
of $\Delta_{c}$ $\sim$ 0.61 eV was obtained from fitting $\rho_{c}$ (not shown).

\begin{figure}
\centering
\includegraphics*[trim = 20mm 20mm 20mm 10mm,width=1\columnwidth,clip=true]{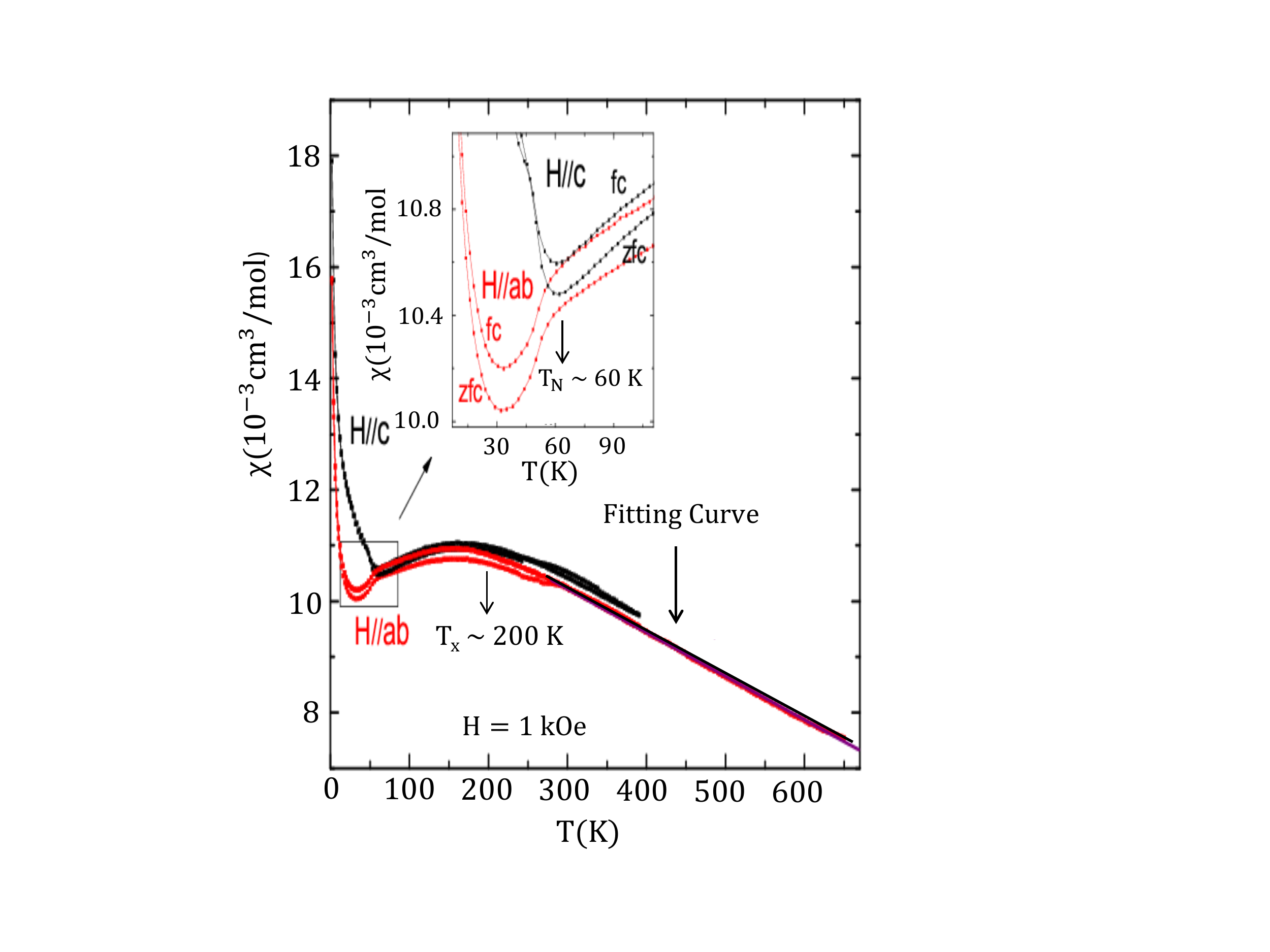}
\caption{(Color online) Temperature dependence of magnetic susceptibility on
the $ab$-plane ($\chi_{ab}$) and along the c-direction ($\chi_{c}$) measured
under both zero-field-cooling (zfc) and field-cooling (fc) conditions. The characteristic
temperature T$_\mathrm{x}$ is indicated. The inset shows $\chi_{ab}$(T) and $\chi_{c}$(T) near T$_\mathrm{N}$ $\sim$ 60 K.}
\label{fig:BaMnSbO_susceptibility}
\end{figure}

The crossover from the activated to the much slower increase in resistivity at
low temperatures indicates that the system has a finite (albeit low) carrier
density as $T$ approaches zero. However, the low-temperature logarithmic behavior of
both $\rho_{ab}$ and $\rho_c$ suggests anomalous temperature dependence of the
scattering rate for the remaining carriers. The magnetic properties shown below (see Fig.~\ref{fig:BaMnSbO_susceptibility})
may offer a clue to the origin of this scattering.

The magnetic susceptibility $\chi$ $=$ $M$/$H$ was measured at an applied magnetic field $H$ = 1 $\mathrm{kOe}$
between 2 and 700 K. Figure~\ref{fig:BaMnSbO_susceptibility} shows the temperature dependence of
the in-plane ($\chi_{ab}$) and the out-of-plane ($\chi_{c}$) susceptibilities in zero-field-cooling (zfc)
and field-cooling (fc) conditions. Several features are worth noticing: (1) there is almost no difference
between $\chi$(H$||$ab) and $\chi$(H$||$c) above $\sim$ 60 K, i.e., the susceptibility
is isotropic; (2) below 60 K $\chi$(H$||$ab) decreases while
$\chi$(H$||$c) increases; and (3) there is a broad bump in $\chi_{ab}$ and $\chi_c$ centered around T $\sim$ T$_\mathrm{x}$. 
These features are reminiscent of another transition-metal oxide $\mathrm{K_2V_3O_8}$. \cite{Lumsden2001}  
While it is a two-dimensional antiferromagnet, its magnetic susceptibility remains isotropic above $T_N$. 
According to \citet{Lumsden2001}, this is the consequence of Heisenberg magnetic interactions.  
The bump seen in the magnetic susceptibility of $\mathrm{K_2V_3O_8}$ is due to competition between the 
antisymmetric Dzyaloshinskii-Moriya (DM) interaction and the symmetric Heisenberg interaction. 
The competition may also result in a canted magnetic structure. For Ba$_\mathrm{2}$Mn$_\mathrm{2}$Sb$_\mathrm{2}$O, 
it is unclear whether the DM interaction should be taken into account. However the increase of $\chi_{c}$ 
below T$_\mathrm{N}$ suggests that spins are canted as well in Ba$_\mathrm{2}$Mn$_\mathrm{2}$Sb$_\mathrm{2}$O, 
which gives rise to  weak ferromagnetism along the c-direction. This is further supported by the non-linear 
field dependence of c-axis magnetization below T$_\mathrm{N}$ (not shown). On the other hand, the characteristic temperature
where $\chi$ shows a bump coincides with the crossover temperature, $T_x$, where the resistivities
($\rho_{ab}$ and $\rho_{c}$) change their temperature dependence. As will be discussed later, this suggests 
that there may be another mechanism to cause the change in both the magnetic susceptibility 
and electrical resistivity.
   
Fitting the data above 350 K with a Curie-Weiss expression
$\chi$ = $\displaystyle \chi_{o} + \frac{C_{CW}}{T-\Theta_{CW}}$
($\chi_{o}$, C$_{CW}$, and $\Theta_{CW}$ are constants), we obtain $\Theta_{CW}$ = -564 K,
C$_{CW}$= 9.24 cm$^{3}$ K/mol, and $\chi_{o}$ = -6.6 $\times$ 10$^{-5}$ cm$^{3}$/mol
(see the solid line in Fig.~\ref{fig:BaMnSbO_susceptibility}).
With a Curie-Weiss constant C$_{CW}$ $=$ $\displaystyle \frac{N_{A} \mu_{eff}^{2}}{3 k_B}$, we estimate the
effective magnetic moment, $\mu_{eff}$ = 2.7 $\mu_{B}$/f.u. 
The negative $\mathrm{\Theta_{CW}}$ suggests antiferromagnetic correlations at
high temperatures leading to long-range antiferromagnetic ordering at T$_\mathrm{N}$ $\sim$ 60 K.
As shown in the inset of Fig.~\ref{fig:BaMnSbO_susceptibility}, $\chi_{ab}$ decreases, indicating
that ordered magnetic moments are in the $a-b$-plane. The fact that $\Theta_{CW}$ is much higher than T$_\mathrm{N}$ suggests
low-dimensional magnetism, because the given crystal structure is not likely to cause magnetic
frustration.

\begin{figure}
\centering
\includegraphics*[trim = 20mm 20mm 20mm 10mm, width=1\columnwidth,clip=true]{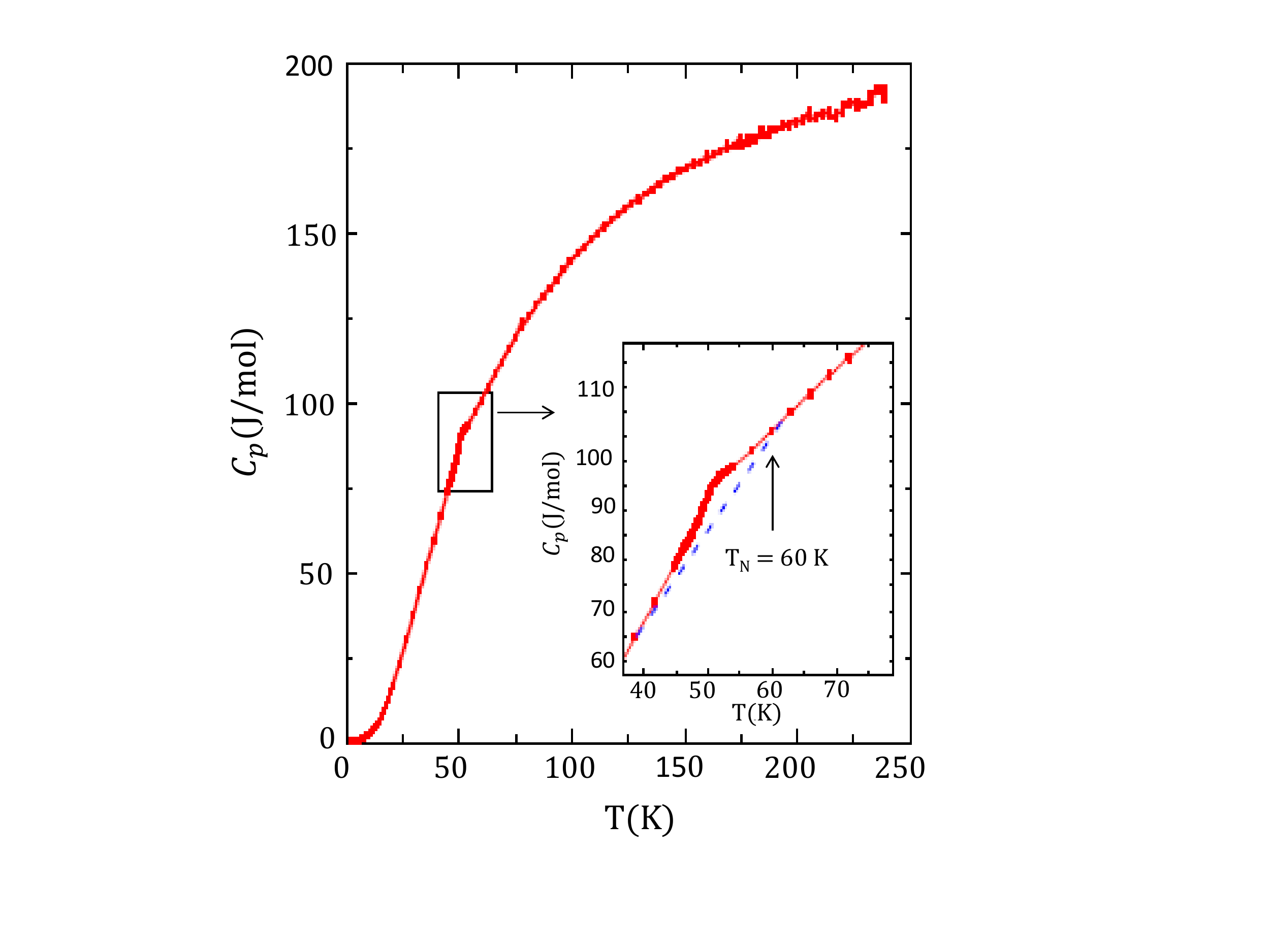}
\caption{(Color online) Temperature dependence of the specific heat, C$_{p}$. The inset shows
the anomaly near T$_\mathrm{N}$, and the dash line represents the background.}
\label{fig:BaMnSbO_specific_heat}
\end{figure}

To confirm the phase transition at T$_\mathrm{N}$, we have measured the specific heat of
Ba$_\mathrm{2}$Mn$_\mathrm{2}$Sb$_\mathrm{2}$O at constant pressure. As shown in  Fig.~\ref{fig:BaMnSbO_specific_heat},
there is clearly an anomaly at T$_\mathrm{N}$, an indication of a phase transition. If we estimate the background
by fitting the data away from T$_\mathrm{N}$, as shown in the inset of Fig.~\ref{fig:BaMnSbO_specific_heat} by the dashed line,
we can obtain the specific heat change $\Delta C_{p}$ due to the phase transition.
By integrating $\displaystyle \int$$\frac{\Delta C_{p}}{T}$$dT$, we find an
entropy change  $\Delta$S $\sim$ 1.55 J/mol-K.
This value is much smaller than the theoretical expectation  S$_{M}$ = $R$ $ln$ $(2S+1)$ $=$
14.9 J/mol-K, ($R$ $=$ 8.314 J/mol-K, with full moment $S$ $=$ 5/2 for Mn$^{2+}$).
This implies that the entropy is either removed prior to the transition and/or
results from a reduced effective magnetic moment.

\begin{figure}
\centering
\includegraphics*[trim = 0mm 14mm 4mm 8mm, clip,totalheight=0.23\textheight, width=3.2in]{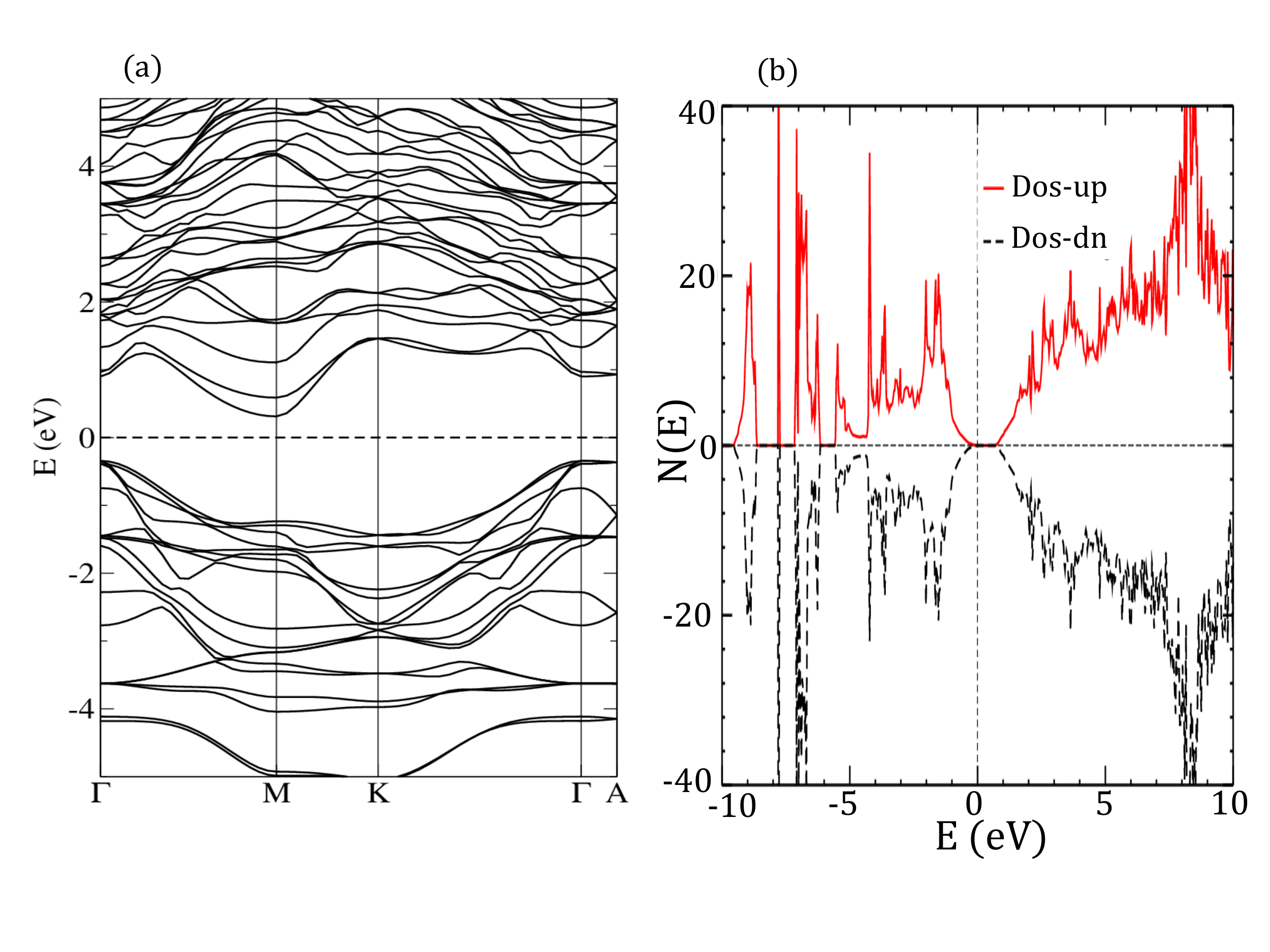}
\caption{(Color online). (a) Calculated band
structure of Ba$_\mathrm{2}$Mn$_\mathrm{2}$Sb$_\mathrm{2}$O. The horizontal dotted line
is the chemical potential, which has been set equal to zero. (b) Calculated
density of states (in units of states/eV-cell) of Ba$_\mathrm{2}$Mn$_\mathrm{2}$Sb$_\mathrm{2}$O.
Both figures have been produced by using the effective Coulomb interaction (U$_{eff}$)
obtained self-consistently at the DFT+U level.}
\label{fig:BaMnSbO_electronic}
\end{figure}

To help understand these experimental results we have performed first-principles ab-initio electronic
structure computations. In order to establish the ground
state of Ba$_\mathrm{2}$Mn$_\mathrm{2}$Sb$_\mathrm{2}$O, we carried out calculations on several configurations:
ferromagnetic, antiferromagnetic, and nonmagnetic.
The results of our computations show that the antiferromagnetic state is the most stable one. 
We used the A-type antiferromagnetic alignment, i.e., the magnetic structure having 
ferromagnetic alignment in-planes but antiferromagnetic coupling between planes. In this structure, no 
supercell is needed. \cite{Sekine1997,Ekuma}
Fig.~\ref{fig:BaMnSbO_electronic}(a) depicts the
band structure of Ba$_\mathrm{2}$Mn$_\mathrm{2}$Sb$_\mathrm{2}$O along the high symmetry points
of the Brillouin zone, while Figure~\ref{fig:BaMnSbO_electronic}(b)
shows the density of states.

The conduction bands  are predominantly Mn 3${d}$ states while the valence bands
are formed by a strong hybridization between Mn 3${d}$ and Sb 5${p}$ with O 2${p}$ at
$\sim$~2.5 eV below the bottom of the conduction band. Also, we find a very large energy difference, 
$E_{Diff}$ $=$ 0.21 eV/Mn, between
the ferromagnetic (the next stable state) and antiferromagnetic (ground state phase) orderings. This indeed is
supported by the large Curie-Weiss temperature ($\mid$$\mathrm{\Theta_{CW}}$$\mid$ $\approx$ 560 K) extracted
from the high-temperature susceptibility.
The maximum of the valence
bands occurs at the $\Gamma$ point and the minimum of the
conduction bands is at the $M$ point. The minimal indirect gap between $M$ and  $\Gamma$ is 0.686 eV,
in close agreement with the experimental values of 0.59 and 0.61 eV deduced
from the conductivity $\sigma_{ab}$ and $\sigma_{c}$, respectively. The effective mass in the 
conduction band is $m^* \sim$ 4.09 $m_o$ (where $m_o$ is
the bare electron mass). In the valence band, the effective mass tensor is rather 
anisotropic but the masses are of the same order. 
We also calculated the magnetic moment
per formula unit. The computed effective magnetic moment $\mu_{eff}$ = 2.52 $\mu_{B}$/f.u., is in
good agreement with the experimental
value of 2.7 $\mu_B$/f.u. obtained from a Curie-Weiss fit to the high temperature susceptibility.
Treating the compound as an intrinsic semiconductor, we obtain a rough estimate of 
the carrier concentration at room temperature
$n_T\sim$ 3.5 $\times$ 10$^{14}$ cm$^{-3}$ using the computed gap, or an order or 
magnitude greater if we take the gap values from the resistivity fit above. 
Simple estimate of the Drude relaxation time using the former value of the carrier density
yields $\tau$ $\sim$ 4.5 $\times$ 10$^{-12}$ s. 
The carrier mobility at room temperature is $\mu$ $=$ $\sigma$/$n$$e$ $\sim$ 2 $\times$ 10$^{3}$ $cm^2$$V^{-1}s^{-1}$.
Residual conductivity at low $T$ indicates that there are mobile carriers that are not 
induced by temperature, but their density is low: even if we take 
the value of the room temperature mobility as a guide, the non-intrinsic 
carrier density $n_i\sim 10^{13}$ cm$^{-3}$ is sufficient to give the 
residual $\rho_{ab}$. In reality, the mobility likely grows by more than an order of 
magnitude as the temperature is lowered, and therefore an even lower carrier density suffices.
Self-consistent structural and geometrical optimization yields lattice parameters:
a = 4.7096 \AA{} and c = 20.0298 \AA{}. These are consistent with the values
determined experimentally. The geometrical optimization reproduced
the atomic positions of Ba$_\mathrm{2}$Mn$_\mathrm{2}$Sb$_\mathrm{2}$O as shown in Table~\ref{table:BaMnSbO_Wyck}.

From the above comparison, we note that the computed results are consistent with those obtained experimentally.
In particular, the computed effective moment is very close to the experimental value
obtained from the Curie-Weiss fit. While Mn$^{2+}$ can be
in high-spin ($S$ = 5/2), intermediate
($S$ = 3/2), or low-spin ($S$ = 1/2) states, it is possible that there is a partial cancellation of the spin by an
antialigned moment from the Sb 5$p$ band of the Sb$_\mathrm{3}$O cage surrounding Mn, as seen in
Yb$_\mathrm{14}$MnSb$_\mathrm{11}$. \cite{Holm2002} According to the X-ray magnetic circular dichroism (XMCD)
measurements, the Sb$_\mathrm{4}$ cage in Yb$_\mathrm{14}$MnSb$_\mathrm{11}$ provides a moment $S=1$ opposite to the Mn
moment \cite{Holm2002} due to hybridization between the Mn 3$d$ and Sb 5$p$ orbitals. \cite{PhysRevB.56.6021}
The strong hybridization of Sb 5$p$ with Mn 3$d$ is confirmed by our computations as
explained above. In our ab-initio calculations, the magnetic moment of Sb is found to be significant with an
antialigned moment of $\sim$ -0.1 $\mu_B$ as
observed in Yb$_\mathrm{14}$MnSb$_\mathrm{11}$. \cite{Holm2002}
If we assume that the induced magnetic moment of the Sb$_\mathrm{3}$O cage is slightly less than $S=1$,
and Mn$^{2+}$ is in the intermediate spin state ($S$ = 3/2), the effective magnetic moment would be close
to our experimental and calculated results. For each MnPn$_\mathrm{4}$ (Pn = P, As, Sb, and Bi), accompanying
the induced magnetic moment, there is a hole per formula unit in the bonding valence
bands. \cite{PhysRevB.65.144414} We anticipate a similar situation in Ba$_\mathrm{2}$Mn$_\mathrm{2}$Sb$_\mathrm{2}$O. 
The effective hole concentration for each
MnSb$_\mathrm{3}$O could be slightly different from that of MnPn$_\mathrm{4}$.
The question is whether these holes interact with
the local magnetic moment provided by Mn$^{2+}$ (i.e., Kondo effect), as seen in Yb$_\mathrm{14}$MnSb$_\mathrm{11}$.
\cite{PhysRevLett.95.046401,PhysRevB.72.205207} The logarithmic temperature dependence of the electrical resistivity
(see Fig.~\ref{fig:BaMnSbO_resistivity}(c)), and the decrease of magnetic susceptibility
(see Fig.~\ref{fig:BaMnSbO_susceptibility})
below $\sim$ 200 K, seem to be consistent with the Kondo picture.\\

\section{Conclusion}
\label{sec:conclusion}
We have studied the structural, electronic, magnetic,
and thermodynamic properties of Ba$_\mathrm{2}$Mn$_\mathrm{2}$Sb$_\mathrm{2}$O single crystals. This compound exhibits
semiconductor behavior with thermally activated electrical conduction at high temperatures, where
the energy gap was found to be $\sim$ 0.59 eV from experiment and $\sim$ 0.686 eV
from band calculations.  The magnetic susceptibility reveals a bump at T$_\mathrm{x}$ $\sim$ 200 K
and an antiferromagnetic transition at
T$_\mathrm{N}$ $\sim$ 60 K. The latter was confirmed by specific heat measurements, indicating a true phase transition.
While the present analysis and
discussion is in no way conclusive, we hope that it will spur interest to further study this material.

\begin{acknowledgments}
\noindent Research is in part supported by the National Science Foundation (NSF) Award Numbers DMR-1002622 (J.L., A.K., and R.J.),
NSF LA-SiGMA EPS-1003897 (C.E.E., M.J., I.V., and J.M.), and NSF DMR-0965009 (S.S.) High performance computational resources are provided by the Louisiana Optical Network Initiative 
(LONI).
\end{acknowledgments}


\end{document}